# Multi-stages Proton Acceleration Booster in Laser Plasma Interaction

S. Kawata[1], D. Sato[1], T. Izumiyama[1], T. Nagashima[1], D. Barada[1], W. M. Wang[2], Q. Kong[3], P. X. Wang[3], Z. M. Sheng[2,4]

[1]Graduate School of Engineering, Utsunomiya University,
7-1-2 Yohtoh, 321-8585 Utsunomiya, Japan
[2]Inst. of Physics, CAS, Beijing 100190, China
[3]Institute of Modern Physics, Fudan University, 200433 Shanghai, China
[4]Department of physics, Shanghai Jiao Tong University, Shanghai 200240, China

**Abstract**

A remarkable ion energy increase is demonstrated by several-stage post-acceleration in a laser plasma interaction. Intense short-pulse laser generates a strong current by high-energy electrons accelerated, when an intense short-pulse laser illuminates a plasma target. The strong electric current creates a strong magnetic field along the high-energy electron current in plasma. During the increase phase of the magnetic field, the longitudinal inductive electric field is induced for the forward ion acceleration by the Faraday law. The inductive acceleration and the target-normal sheath acceleration in the multi stages provide a unique controllability of the ion energy. By the four-stage successive acceleration, our 2.5-dimensional particle-in-cell simulations demonstrate a remarkable increase in ion energy by a few hundreds of MeV; the maximum proton energy reaches 254MeV.

**I. INTRODUCTION**

By chirped pulse amplification, a higher laser intensity has been realized, and high intensity short pulse lasers are now available for experiments and applications. On the other hand, ion beams are useful for basic particle physics, medical ion therapy, controlled nuclear fusion, high-energy sources, and so on [1-10]. The energy of ions, which are accelerated in an interaction between an intense laser pulse and the near-critical density target, are over a few tens of MeV [11-29]. The issues in the laser ion acceleration include an ion beam collimation[12, 13], ion energy spectrum control[14], ion production efficiency[28], etc[15]. Depending on ion beam applications, the ion particle energy should be controlled. For example, ion beam cancer therapy needs a few hundred of MeV for proton energy. In recent research works ions are accelerated in an interaction of an ultraintense laser plus with a near-critical-density plasma. One of researchers achieved a few hundred of MeV of ion beam energy[19] with laser intensity $>10^{22}$ W/cm$^2$, but it is difficult to reach the very high laser intensity practically in the present technology. In this paper we focus on a boost of ion particle energy by post-accelerations in a laser plasma interaction. We succeeded to achieve a few hundreds of MeV of the proton beam energy by several times of the ion accelerations in the laser-plasma interaction. In our study, we employ an intense short-pulse laser and the near-critical density plasma target, which consists of hydrogen. Figure 1(a) shows the conceptual diagram for the post-acceleration in the laser plasma interaction. In this paper, we perform 2.5-dimensional particle-in-cell simulations to investigate the ion beam post-acceleration. When the intense laser pulse propagates through the plasma, it accelerates a part of electrons. The electrons form a high current and generate the azimuthal magnetic field. In the laser plasma interaction, the ion dynamics is affected directly by the electric field and the behavior of the electrons[16-18]. The electrons form the strong magnetic field, and during the increase in the azimuthal magnetic field the inductive strong electric field is generated by the Faraday low. The ions are accelerated by the inductive electric field.

## II. MULTI-STAGE ION ACCELERATION

In the 1st acceleration stage, at first the protons are accelerated by TNSA (Target Normal Sheath Acceleration) mechanism[13, 19, 20] in the linear density gradient behind the target surface. Secondly, the high energy ions are selected, and the accelerated ions are sent to the next 2nd post-acceleration. The ion 3rd and 4th post-accelerations are continued. In the 1st and 2nd acceleration stages, the ion beam is mainly accelerated by the TNSA acceleration mechanism. However, in the 3rd and 4th post-accelerations the ions are accelerated by the inductive acceleration based on the Faraday law inside of the plasma target as well as the TNSA mechanism behind the target surface. When an intense short pulse laser illuminates the near-critical density plasma, the inductive acceleration field moves with a speed $v_g$, which is less than $c$ depending on the plasma density: $v_g = c\sqrt{1 - \omega_{pe}^2/\omega^2}$, where $c$ is the speed of light, $\omega_{pe}$ the electron plasma frequency and $\omega$ the laser frequency. Additionally, the beam protons have a higher speed especially in the 4th stage, compared with that in the 1st and 2nd stages. Therefore, the accelerated ions are kept accelerating for a long time inside of the near-critical density plasma target.

In this paper, we perform 2.5-dimensional particle-in-cell simulations. The simulation is divided into 4 stages in order to keep the computing CPU time reasonable. One target is located in one stage, in which a laser accelerates ions. The ions accelerated are loaded to the next simulation stage. This simulation procedure is performed to simulate the multi stage acceleration. The hydrogen plasma target is located in $9.5\lambda < X < 29.5\lambda$ and $11.0\lambda < Y < 39.0\lambda$ in each simulation box. The target density is $0.7n_c$, and the edge region density has a linear gradient scale from $0n_c$ to the maximum density of $0.7n_c$ with the linear density gradient of $2\lambda$ in the $X$ and $Y$ directions at the target edges. Here, $n_c = m\varepsilon_0\omega^2/e^2$ is the critical density, $m$ is electron mass, $e$ is electric charge and $\varepsilon_0$ is dielectric constant in vacuum, respectively. The target model and the lasers employed are identical for all the 4 acceleration stages. For the staging from one stage to the next, the beam ions accelerated are extracted from the right boundary of the simulation box and then transferred to the left boundary of the next simulation box, so that the beam loading is smooth to the next stage. The ion beam transport is consistently computed between the two stages: for example, the ion beam divergence and the beam ion self-charge interaction are naturally included during the beam transport between the stages, although the simulation is divided into the 4 stages. The distance between the two adjacent targets is $60\lambda$, and is so short that additional ion beam bunching and focusing elements are not required. In another view, it may be also reasonable to think the four-stage targets employed in this paper as one longer target with gaps of $60\lambda$. In our simulations, for example, the first laser pulse interacts with the first target, and then the target debris are well expanded and blown off so that the next laser pulse impinging from the left area can reach the next target safely without a significant influence of the former target debris. In addition, after the laser pulse interacts with the target, the laser pulse is depleted well in the target so that the second~fourth targets survives successfully without the illumination of the former laser pulses. The target longitudinal length is selected to deplete the laser pulse energy sufficiently inside the target. The laser intensity is $I=1.0 \times 10^{20}$ W/cm$^2$. The laser spot diameter is $10.0\lambda$, each laser is focused on the left edge of each target, and the pulse duration is 40fs. The laser transverse profile is in the Gaussian distribution, and the laser temporal profile is also Gaussian. The laser wavelength is $\lambda=1.053\mu$m. The simulation box is $80\lambda$ in the longitudinal direction and $50\lambda$ in the transverse direction. The mesh size is selected to be the same order of the Debye length in our simulations, so that the artificial numerical unphysical particle collisions are suppressed in particle-in-cell simulations. The free boundary condition is employed so that the boundaries do not reflect particles and waves. In the following sections we present simulation results and discussions on the relating physics, and the result for the dramatic increase in the beam proton energy by the multi-stage phased ion acceleration.

We study the increase in the ion beam energy from the 1st ion acceleration to the 4th ion post-acceleration. Ion particle energy is one of the issues of the laser plasma ion acceleration. Therefore, in this paper we focus on the boost of ion particle maximal energy by the post-accelerations in the laser plasma interaction. Figures 1(b)-(e) present the longitudinal electric fields at (b)$t=$130fs and (c)190fs, and the magnetic fields at (d)$t=$130fs and (e)190fs for the first stage. The laser generates the high-energy electrons inside of the target. A magnetic field is also formed along the channel in the laser plasma interaction as shown in Figs. 1(d) and (e), and the magnetic vortex is formed at the end of the plasma target[19] (see Fig. 1(e)). During the increase in the magnetic field the inductive electric field is generated as shown in Fig. 1(b) and (c). The phenomenon shown in Figs. 1(b)-(e) is almost identical to that in the 2nd~4th stages, because the identical target and laser are employed in the present work. The fast electrons also induce the acceleration electric field at the behind region of plasma target (see Fig. 1(c)). Therefore, the ions are accelerated by the TNSA acceleration field and also by the inductive electric field. The maximum acceleration electric field reaches to 16.7MV/μm at the end of the target area, and the maximum magnetic field reaches to 37.0kT in the 1st acceleration.

**III. REMARKABLE ION ENERGY INCREASE BY MULTI-STAGE ACCELERATION**

Figure 2 shows the histories of the maximum proton energy from the 1st acceleration to the 4th post-acceleration. The maximum proton energy is 38.9MeV in the 1st acceleration at 700fs, 89.9MeV in the 2nd post-acceleration at 600fs, 149MeV in the 3rd post-acceleration at 500fs. Finally the maximum proton energy reaches 254MeV in the 4th ion post-acceleration at 450fs. The accelerated protons in the 1st ion acceleration are selected, and connected to the next simulation of the 2nd ion post-acceleration. This procedure is successively performed for all the post-acceleration stages. In the 1st acceleration stage, the energy efficiency is 2.12% from the laser to the accelerated (>20MeV) ions. In the 2nd post-acceleration, the 17.4% of the input beam ions are accelerated over 60MeV from the input ion beam. The energy efficiency form the laser to the accelerated ion is 0.566%. In the 3rd post-acceleration, the accelerated ions over 120MeV are 7.58% by input beam ions. The energy efficiency form the laser to the accelerated ion is $4.67 \times 10^{-2}$ %. In the 4thd post-acceleration, the accelerated ions over 200MeV over are 5.87% by input ions. The energy efficiency form the laser to accelerated ion is $1.46 \times 10^{-2}$ %. In the 2nd~4th post-acceleration stages of course lower-energy ions are also accelerated as in the 1st stage, but the lower-energy (~20MeV) ions are not included in the energy efficiency. The maximum proton energy is remarkably accelerated almost twice in the 3rd and 4th post-accelerations. In the case of the 1st and 2nd acceleration, the protons are accelerated almost exclusively by the TNSA mechanism. However, in the 3rd and 4th post-accelerations, first the ions are accelerated by the inductive field inside the plasma body, and then accelerated further by the TNSA. In the 3rd and 4th accelerations, when an intense short pulse laser illuminates the near-critical density plasma, the inductive acceleration field moves with a speed less than *c*, depending on the plasma density. Especially, in the 4th stage the proton energy is about 150~250MeV during the inductive acceleration phase (Fig. 2), and therefore the proton speed is about $0.566c \sim 0.653c$, which corresponds to $v_g \sim 0.548c$ (the speed of laser in the plasma density of $0.7n_c$). Therefore, the ions are accelerated continuously by the inductive field in the target and then accelerated further by the TNSA field successfully.

Figure 3(a) shows the acceleration electric field along the center line of target in the longitudinal direction. The black solid sine is $Ex$ at $t=90$fs and the grey dotted line is $Ex$ at $t=160$fs for the 4th post-acceleration in Fig. 3(a). The inductive electric field for the ion acceleration is generated inside of the near-critical density plasma target (Fig. 3(a) solid line). The acceleration electric field for TNSA is generated behind the target (Fig. 3(a) dotted line). Figure 3(b) shows the proton spatial distributions and Fig. 3(c) shows the proton energy distributions in $20.0\lambda < Y < 30.0\lambda$ at $t=0$fs, 90fs, 160fs, 240fs and 450fs. The proton energy reaches to 254MeV at $t=450$fs. In this paper the laser spot size is $10\lambda$, and as shown in Figs. 3(b) and 1(b) the core protons accelerated are almost located in $20.0\lambda < Y < 30.0\lambda$.

Figures 4(a)-(d) show the ion beam energy spectra from the 1st acceleration to the 4th post-acceleration. The energy spectra of protons in (a) the 1st acceleration, in (b) the 2nd ion post-acceleration at $t=0$fs and 600fs, in (c) the 3rd post-acceleration at $t=0$fs and 500fs and in (d) the 4th post-acceleration at $t=0$fs and 400fs. The dotted lines show that for the input ion beam for the 2nd - 4th stages, and the solid lines are the outputs. Figures 4 present all the ions, including the scattered ions transversely (see Fig. 3(b)). The inset figures in Fig. 4 show the high-energy core part of the beam ions located in $20.0\lambda < Y < 30.0\lambda$, and the core part of the beam ions could be useful for practical purposes.

In the present case, our simulation parameter study shows that the synchronization timing requirement $\Delta t$ between the laser pulse and the ion beam is about 10 fs so that the decrease in the ion maximal energy is limited by 10.2%. The TNSA field and the inductive acceleration field are generated by the laser pulse, whose pulse duration time is 40fs in this paper. The laser pulse risetime $T_{Lr}$ is ~20fs. Especially the risetime $T_{Lr}$ defines the duration time for the inductive acceleration field in the plasmas. Therefore, the synchronization requirement should be less than the laser risetime: $\Delta t << T_{Lr}$.

## IV. CONCLUSIONS

In this paper, we investigated and clarified the generation of high energy ion beam by the multi-stage acceleration. In our study, we succeeded to reach to the ion beam energy of about 254MeV by the four-stages ion acceleration in the near-critical density plasma target. In the 4th post-acceleration, the energy spectrum has the peak around 207MeV. The multi-stage ion beam post-acceleration by the near-critical density plasma target may serve a new way to create high energy ion beams in the future laser ion accelerator with the controllability of ion energy. In this paper, we focus on the issue of ion energy increase and control in laser accelerator even for a small number of ions. Therefore, we do not focus on the energy efficiency; the overall energy efficiency in the specific case shown in the paper is quite low as shown above, because the identical laser pulses are employed for all the acceleration stages for simplicity and in general the energy efficiency in the post-acceleration stages tends to be low. The first stage may provide a relatively high energy efficiency as presented above. When a higher energy laser is employed at the first or beginning stage, the overall energy efficiency is defined by the stage in which a larger laser energy is used, and the overall energy efficiency may be kept being a reasonable value. The studies on tthe energy efficiency, the laser prepulse and also the ion energy spectrum control should be addressed in the next near-future work.

**ACKNOWLEDGEMENTS**

This work is partly supported by MEXT, JSPS, KEK, ILE/Osaka Univ. and CORE (Center for Optical Research and Education, Utsunomiya Univ., Japan).

**Figure Captions**

Fig. 1 (a) The conceptual diagram for the post-acceleration in the laser plasma interaction. The generated ions from laser ion source are accelerated by several-stages post-acceleration. We employ the near-critical density plasma target, which consists of hydrogen. In this study, the four-stages ion acceleration is performed. The longitudinal electric fields, which contribute the ion acceleration at (b)$t$=130fs and (c)190fs, and the magnetic fields at (d)$t$=130fs and (e)190fs. The laser generates the high-energy electron inside of the target. A transverse magnetic field is also formed along the channel in the laser plasma interaction. During the increase phase of the magnetic field the inductive longitudinal electric field is created. The fast electrons also induce the acceleration electric field at the behind region of plasma target, so that the ions are accelerated also by the TNSA acceleration mechanism.

Fig. 2 The histories of the maximum proton energy from the 1st acceleration to the 4th post-acceleration. The maximum proton energy is remarkably accelerated by the four-stages acceleration. The maximum proton energy is finally about 254.0MeV in the 4th post-acceleration at 450fs.

Fig. 3 (a) Acceleration electric fields averaged over the laser one cycle along the center line of the plasma target in the longitudinal direction. The black solid line is $E$x at $t$=90fs and the grey dotted line is at $t$=160fs in the 4th post-acceleration. (b) The proton spatial distributions at $t$=0fs, 90fs, 160fs, 240fs and 450fs. The color shows the energy of the protons. (c) The energy distributions at $t$=0fs, 90fs, 160fs, 240fs and 450fs for the protons existing in 20.0λ < $Y$ < 30.0λ at 450fs in Fig. 3(b). The maximum proton energy reaches to finally 254MeV in the 4th post-acceleration at 450fs.

Fig. 4  Energy spectra of protons in (a) the 1st acceleration, in (b) the 2nd ion post-acceleration at $t$=0fs and 600fs, in (c) the 3rd post-acceleration at $t$=0fs and 500fs and in (d) the 4th post-acceleration at $t$=0fs and 400fs. The dotted lines are the input the ion beams to the next, and the solid lines are the output ion beams from each acceleration stage. Figures 4 present all the ions, including the scattered ions transversely (see Fig. 3(b)), and the inset figures in Fig. 4 show the high-energy core part of the beam ions located in 20.0λ < $Y$ < 30.0λ. The core part of the beam ions could be useful for practical purposes.

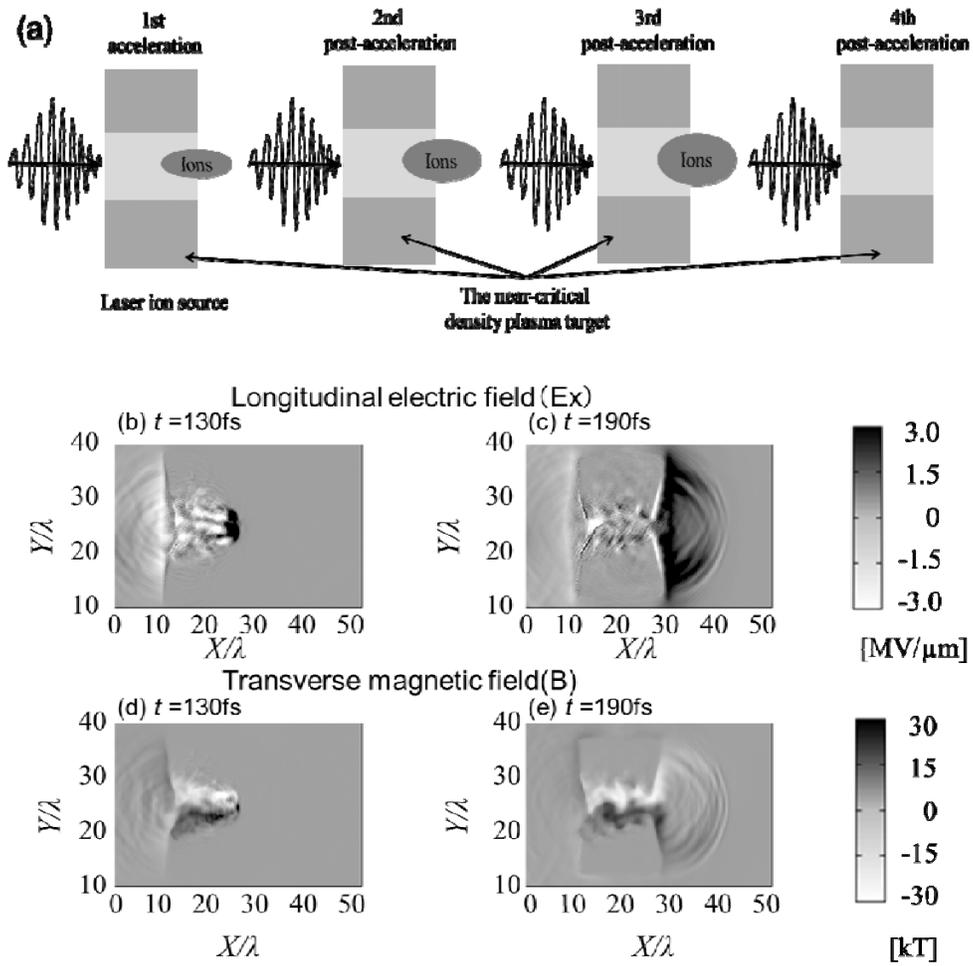

Figure 1 S. Kawata, *et al.*

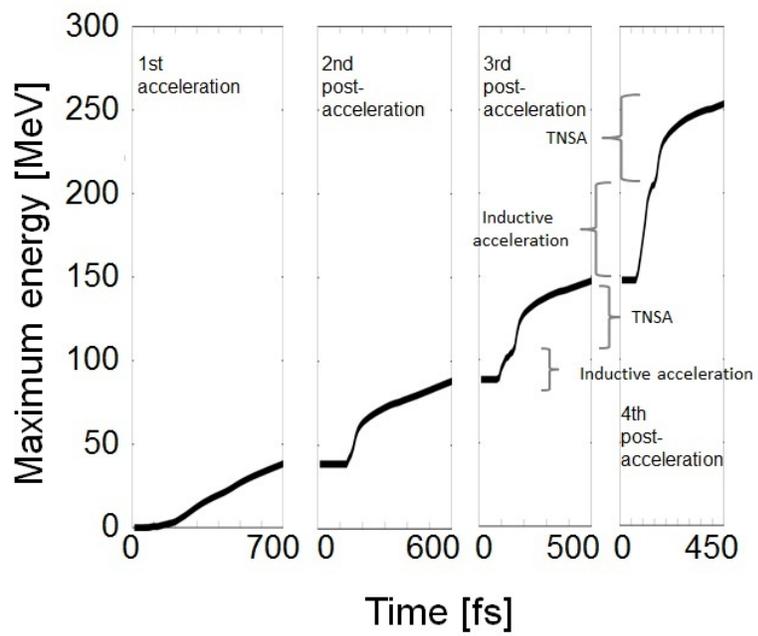

Figure 2 S. Kawata, *et al.*

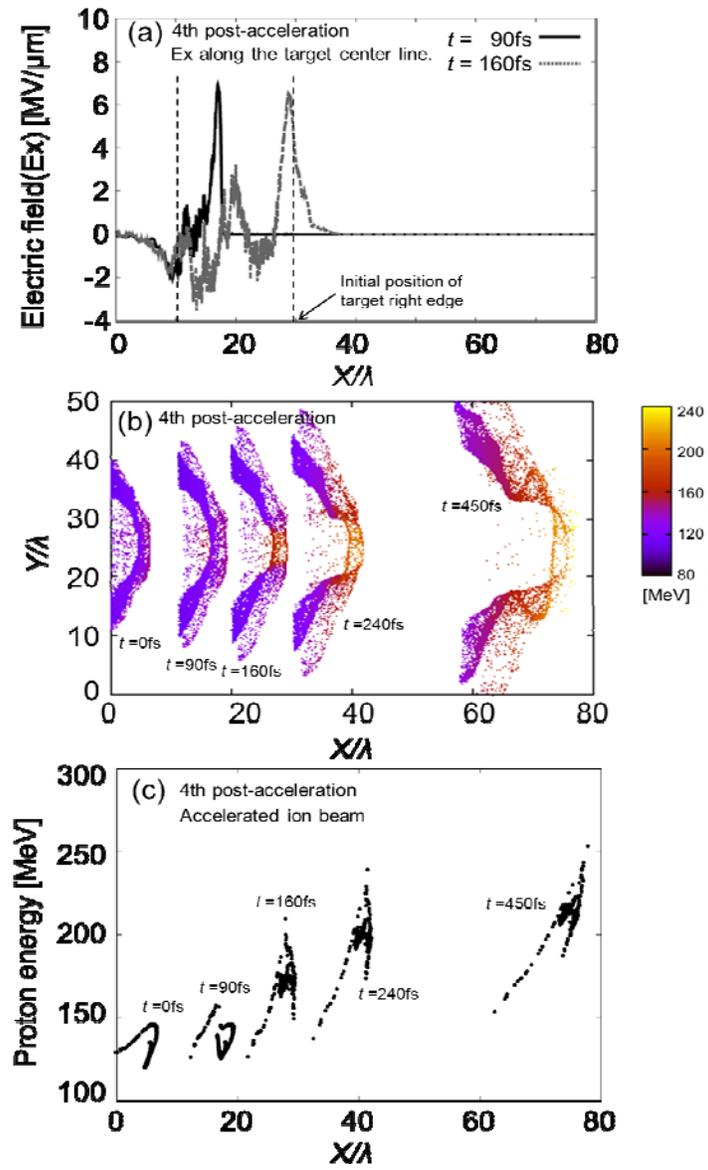

Figure 3 S. Kawata, *et al.*

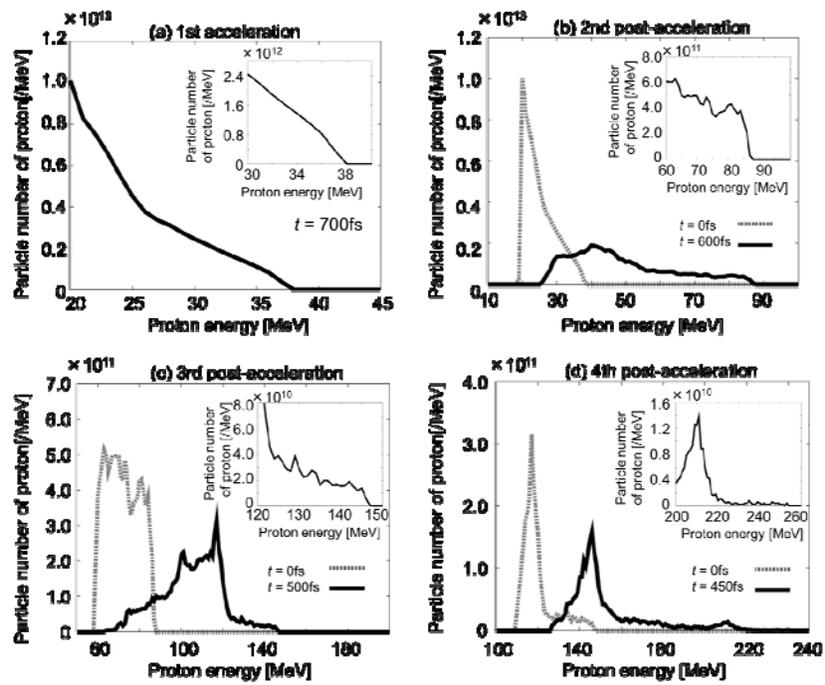

Figure 4 S. Kawata, *et al.*